\documentclass[sigplan,10pt]{acmart}
\usepackage{booktabs}   
\usepackage{subcaption} 

\usepackage{mathsprograms}

\usepackage[flushleft]{threeparttable}
\usepackage{multirow}
\usepackage{makecell}
\usepackage[outdir=./]{epstopdf}

\usepackage{enumitem}

\usepackage{syntax}
\usepackage{amsmath}
\usepackage{mathpartir}
\usepackage{algorithm}
\usepackage{algpseudocode}
\usepackage{stmaryrd}
\usepackage{tikz}
\usepackage{array}
\usepackage{hyperref}
\usetikzlibrary{shapes.geometric, arrows}

\usepackage{listings, xcolor}

\definecolor{verylightgray}{rgb}{.97,.97,.97}

\newcommand{\lstbg}[3][0pt]{{\fboxsep#1\colorbox{#2}{\strut #3}}}

\lstdefinelanguage{Solidity}{
	keywords=[1]{anonymous, assembly, assert, balance, break, call, callcode, case, catch, class, constant, continue, constructor, contract, debugger, default, delegatecall, delete, do, else, emit, event, experimental, export, external, false, finally, for, function, gas, if, implements, import, in, indexed, instanceof, interface, internal, is, length, library, log0, log1, log2, log3, log4, memory, modifier, new, payable, pragma, private, protected, public, pure, push, require, return, returns, revert, selfdestruct, send, solidity, storage, struct, suicide, super, switch, then, this, throw, transfer, true, try, typeof, using, value, view, while, with, addmod, ecrecover, keccak256, mulmod, ripemd160, sha256, sha3, handler, bind}, 
	keywordstyle=[1]\color{blue}\bfseries,
	keywords=[2]{address, bool, byte, bytes, bytes1, bytes2, bytes3, bytes4, bytes5, bytes6, bytes7, bytes8, bytes9, bytes10, bytes11, bytes12, bytes13, bytes14, bytes15, bytes16, bytes17, bytes18, bytes19, bytes20, bytes21, bytes22, bytes23, bytes24, bytes25, bytes26, bytes27, bytes28, bytes29, bytes30, bytes31, bytes32, enum, int, int8, int16, int24, int32, int40, int48, int56, int64, int72, int80, int88, int96, int104, int112, int120, int128, int136, int144, int152, int160, int168, int176, int184, int192, int200, int208, int216, int224, int232, int240, int248, int256, mapping, string, uint, uint8, uint16, uint24, uint32, uint40, uint48, uint56, uint64, uint72, uint80, uint88, uint96, uint104, uint112, uint120, uint128, uint136, uint144, uint152, uint160, uint168, uint176, uint184, uint192, uint200, uint208, uint216, uint224, uint232, uint240, uint248, uint256, var, void, ether, finney, szabo, wei, days, hours, minutes, seconds, weeks, years, signal},	
	keywordstyle=[2]\color{teal}\bfseries,
	keywords=[3]{block, blockhash, coinbase, difficulty, gaslimit, number, timestamp, msg, data, gas, sender, sig, value, now, tx, gasprice, origin},	
	keywordstyle=[3]\color{violet}\bfseries,
	identifierstyle=\color{black},
	sensitive=true,
	comment=[l]{//},
	morecomment=[s]{/*}{*/},
	commentstyle=\color{gray}\ttfamily,
	stringstyle=\color{red}\ttfamily,
	morestring=[b]',
	morestring=[b]",
	morecomment=[f][\lstbg{red!20}]-,
	morecomment=[f][\lstbg{green!20}]+,
  	morecomment=[f][\textit]{@@}
}

\definecolor{codegreen}{rgb}{0,0.6,0}
\definecolor{codegray}{rgb}{0.5,0.5,0.5}
\definecolor{codepurple}{rgb}{0.58,0,0.82}
\definecolor{backcolour}{rgb}{0.95,0.95,0.92}
\newcommand{\llll}{\mathrel{{=}{\llbracket}}}
\newcommand{\rrrr}{\mathrel{{\rrbracket}{\Rightarrow}}}

\AtBeginDocument{%
  \providecommand\BibTeX{{%
    \normalfont B\kern-0.5em{\scshape i\kern-0.25em b}\kern-0.8em\TeX}}}

\setcopyright{none}
\renewcommand\footnotetextcopyrightpermission[1]{}
\newcommand*{\name}{{SigVM}}
\newcommand*{\namelang}{{SigSolid}}
\newcommand*{\namechain}{{SigChain}}
\newcommand{\sourcecode}[1]{\texttt{#1}}

\begin{document}

\title[Short Title]{SigVM: Enabling Event-Driven Execution for Autonomous Smart Contracts}         

\author{Zihan Zhao}

\affiliation{
  \institution{University of Toronto}            
}
\email{simon.zhao@mail.utoronto.ca}          
\author{Sidi Mohamed Beillahi}

\affiliation{
  \institution{University of Toronto}            
}
\email{sm.beillahi@utoronto.ca}          
\author{Ryan Song}

\affiliation{
  \institution{University of Toronto}            
}
\email{r.song@mail.utoronto.ca}          
\author{Yuxi Cai}

\affiliation{
  \institution{University of Toronto}            
}
\email{yuxijune.cai@mail.utoronto.ca}          
\author{Andreas Veneris}

\affiliation{
  \institution{University of Toronto}            
}
\email{veneris@eecg.utoronto.ca}          
\author{Fan Long}

\affiliation{
  \institution{University of Toronto}            
}
\email{fanl@cs.toronto.edu}          
\begin{abstract}
This paper presents {\name}, a novel blockchain virtual machine that supports
    an event-driven execution model, enabling developers to build autonomous smart contracts. 
    Contracts in {\name} can emit signal events, on
    which other contracts can listen. Once an event is triggered, corresponding
    handler functions are automatically executed as signal transactions. We
    build an end-to-end blockchain platform {\namechain} and a contract
    language compiler {\namelang} to realize the potential of {\name}.
    Experimental results show that our benchmark applications can be
    reimplemented with {\name} in an autonomous way, eliminating the
    dependency on unreliable mechanisms like off-chain relay servers. The
    development effort of reimplementing these contracts with {\name} is small, i.e., we
    modified on average $2.6\%$ of the contract code. 
\end{abstract}




\maketitle

\vspace{-3mm}
\section{Introduction}

Blockchain has become a revolutionary technology that powers decentralized
ledgers at internet-scale. Ethereum, the second largest blockchain, introduces
smart contracts, which further fuel blockchain innovations on real-world
applications in various domains, including financial systems, supply chains,
and health cares. A smart contract is a program operating on the blockchain
ledger to encode customized transaction rules. Once deployed, the contract and
the encoded rules are then faithfully executed and enforced by all participants
of the blockchain platform, eliminating any potential counter-party risks in
the future. 


Smart contract in Ethereum can generate \emph{events}, each of which is a
developer-defined record of data to log state changes or contract activities
that are important to external observers. To implement events, Ethereum Virtual
Machine (EVM)~\cite{buterin2014ethereum} puts all generated event data into a dedicated region of the
blockchain state called \emph{event logs}. This region is \emph{write-only} for
smart contracts but it can be queried by any external user who runs an Ethereum
full node. The original design of the event mechanism in Ethereum is to
facilitate the integration of on-chain components and off-chain components in a
blockchain-powered application. 
For example, the transfer function of the popular ERC-20 contract for fungible tokens typically emits a transfer event besides
updating the token ledger state in the contract~\cite{erc-20}. It expects that a digital
wallet application will run an Ethereum full node as its back-end, monitor
these events, and update its front-end GUI accordingly to show the token balance to users. 

As smart contracts become more and more complicated and
inter-dependent, the existing event primitive in Ethereum
becomes increasingly inadequate. In many scenarios, the correctness of one smart
contract is now dependent on its timely responses to critical events from other
smart contracts. For example, in Ethereum, there are oracle
contracts~\cite{provable, chainlink}, which feed off-chain data, such as the digital asset prices, to the
blockchain; there are also decentralized
finance (DeFi) smart contracts~\cite{makerdaowp, compoundwp, airswapwp}, for
which the most recent price of a digital asset is critical, e.g., collateral
liquidation is required when the asset price drops below a certain threshold. The DeFi contracts therefore have to timely respond to asset price change events from the oracle contracts. Unfortunately, it is impossible to implement such an
event-driven execution model in Ethereum. This is because, 1) event logs are write-only
for smart contracts and a contract cannot respond to emitted events from other contracts automatically; 
2) a smart contract execution can only be triggered (in)directly via function calls by external user transactions.

To this end, many blockchain applications circumvent this problem via
\emph{off-chain relay servers}~\cite{relayer}. A server constantly monitors
the blockchain ledger. When a critical event from a source contract (e.g., an
oracle contract) occurs, the server will send a \emph{poke transaction} to a
target contract (e.g., a DeFi contract) to drive the contract to respond to the event. However,
this adhoc solution has two undesirable consequences. First, it creates a
central point of failure that defeats the purpose of encoding transaction rules
as smart contracts on blockchain platforms. If the off-chain relay server of the
target contract fails, the contract will not properly respond to 
critical events. Secondly, the blockchain platform may not process the poke transaction timely due to insufficient transaction fees or network congestion. The target contract may undesirably interact with other users 
before it incorporates the critical updates carried by the poke transaction.


In this paper, we present a novel end-to-end blockchain platform that extends
Ethereum to support an event-driven execution model. The core of our framework is
{\name}, a novel virtual machine that extends EVM with new opcodes to introduce
\emph{signal events}, a special kind of events that enables a new way for
multiple contracts to interact with each other. To realize {\name}, we develop
{\namelang}, a modified version of Solidity programming language that can
utilize new opcodes in {\name}, and {\namechain}, a prototype blockchain
platform that implements {\name}. 

In {\name}, contracts can emit signal events, on which other contracts can
listen. When a contract listens to a signal event, it binds a function as its
handler. 
When the event is triggered, the handler function will be automatically executed. 
This new signal event mechanism enables autonomous smart contracts to timely
respond to critical events, eliminating the reliance on off-chain relay servers.

One challenge {\name} faces is how to integrate the execution of handler
functions with the existing smart contract framework. In {\name}, a transaction
from a user to a smart contract can emit signal events and it may therefore
trigger multiple handler functions. If we naively implement the execution of
these functions synchronously as function calls, the cascading execution
process may cause the transaction to exceed the block gas limit. Furthermore, a
user who triggers a signal event will have to pay the gas cost for the
execution of all of the associated handler functions. This is undesirable,
because the user triggering the event is often the service provider while the
contracts containing the handler functions are service users. For example, an
oracle maintainer sends a transaction to trigger an event to update the price
of a digital asset in an oracle contract and this event is listened by many
other contracts to react upon the price update. It is counter-intuitive to ask the
oracle maintainer to pay for the execution cost of the contracts using the
oracle service.

To address this challenge, {\name} executes handler functions asynchronously as
\emph{signal transactions}, a special kind of transactions generated by the
{\name} execution engine when a signal event is emitted. These special
transactions will be packed along with regular transactions in blocks. This
asynchronous execution mechanism enables {\name} to circumvent the block gas
limit issue. Because {\name} executes signal transactions asynchronously,
miners can pack them in separate blocks. This mechanism also enables {\name} to
charge the transaction fee of each signal transaction differently to 
introduce proper incentives. 
The smart contract that binds a handler function pays the transaction fee of
the corresponding signal transaction in {\name}. The transaction fee will be
slightly higher than the average of normal transactions in the same
block\footnote{The average value is computed by miners during transactions
packing.}, which incentivizes miners to prioritize their execution. 



Another challenge {\name} faces is the possibility that a smart contract
interacts with other users undesirably before it can integrate critical state
updates in an asynchronous signal transactions. Note that the centralized
off-chain relay server solution faces the same challenge. The typical solution
of paying high transaction fee does not eliminate this risk, because powerful
miners can selectively pack hijacking transactions ahead of the signal/poking
transactions (i.e., \emph{hijacking manipulations}).

{\name} addresses this challenge with a novel \emph{contract event lock}
mechanism. If a contract has any pending signal transactions, {\name} allows to
lock the contract. Any normal transactions interacting with a locked contract
become no-ops. The miner who packs the transaction will receive no transaction
fees and the transaction will be recycled back to pending transaction pools. It
may be packed again in the future when the pending signal transactions are
processed and the contract is unlocked. This mechanism effectively stops any
hijacking transactions from exploiting a contract in the middle of signal event
handling.


We evaluate {\name} by reimplementing $23$ smart contracts from $13$ popular
decentralized applications that are critical to the economic ecosystem of
Ethereum. Our results show that the signal event mechanism in {\name}
efficiently replaces off-chain relay servers and reduces the applications'
vulnerability to hijacking manipulations. Our results also show that the
{\namelang} language powered by {\name} is practical to use. The development
effort of migrating these contracts to {\name} is small, i.e., we modified on
average $2.6\%$ of the contract code. 


In summary, this paper makes the following contributions:
\begin{itemize}
\item \textbf{{\name}: } a novel blockchain virtual machine that extends EVM with an event-driven execution model to enable autonomous smart contracts (\S\ref{sec:design}-\ref{sec:implementation}). 

\item \textbf{Signal transaction and contract event locking: } a novel asynchronous signal transaction design to address the series of challenges of integrating {\name} with a blockchain platform (\S\ref{sec:design}-\ref{sec:implementation}). A novel contract event locking mechanism to prevent hijacking manipulation against signal transactions (\S\ref{sec:design}-\ref{sec:implementation}).

\item \textbf{Implementation:} an end-to-end prototype blockchain platform powered by {\name} (\S\ref{sec:implementation}). 
\item \textbf{Experimental evaluation:}  an evaluation using $23$ smart contracts from $13$ popular distributed applications (\S\ref{sec:results}). Our results show that {\name} is practical and it
    enables the development of significantly more autonomous and robust contracts. 
\end{itemize}

Aside from the aforementioned technical sections, we present a motivating example of developing smart contracts in {\name} in Section~\ref{sec:example}, and discuss related work and conclusions in Sections~\ref{sec:relatedwork} and~\ref{sec:conclusion}.


\section{Example}
\label{sec:example}
Listing~\ref{lst:maker} presents the simplified code snippet of parts of MakerDAO to illustrate building autonomous contracts with {\name}. 
Lines preceded with "-" and "+" are changes we made to the original implementation to adopt {\namelang}. MakerDAO is a decentralized finance protocol,
which provides collateral-backed stablecoin called Dai~\cite{makerdaowp}. It
follows the Maker Protocol to keep Dai softly pegged to USD by a fixed ratio. This is accomplished by backing Dai with crypto assets based on the market
prices. In order to generate Dai, users send collaterals to the Collateralized 
Debt Position smart contract to create a vault. To meet the long-term solvency
of the system, the ratio between the deposit to the vault and the Dai the vault
owner obtains must be greater than a threshold decided by a MakerDAO governance
committee at all time. Once the ratio drops below the set threshold, the Maker Protocol forces the collateral to be liquidated. 

\begin{lstlisting}[language=Solidity,caption={Simplied Code Snippet of MakerDAO}, label={lst:maker}, ,escapechar=|]
contract Median {
  uint val; // price
  //Peek to get current price of a digital asset
  function peek() public view returns (uint,bool) |\label{line:pk}|
   { return (val, val > 0); }
+ signal Pr(uint,bool); //price update signal |\label{line:signal}|
+ address[] SigT; //Target handler addresses
  //Feed to set current price of a digital asset
  function feed(uint[] p,uint[] v,bytes[] r,bytes[] s, address[] _sigT) public { |\label{line:fd}|
   // check the signature info in r, s, v
   ...
   val = computeMedian(p);
+  SigT = _sigT; // targets are only OSM modules
+  Pr.emit(val,val>0).target(SigT).delay(0);//emit to OSM|\label{line:emit}|
  } ... }
contract OSM {
  struct Feed { uint val; bool has; }
  Feed cur, nxt; //obtained prices
  uint b; //last price feed block number
+ address[] SigR; //allowed accounts |\label{line:SigR1}|
+ bytes[] SigM; //accessible methods |\label{line:SigM1}|
  //Spotter call peek to acquire price feed
- function peek() public view returns(uint,bool)
-    { return (cur.val, cur.has); }
+ signal DPr(uint,address) //delayed price update signal |\label{line:signalosm}|
+ address[] SigT; //target handler addresses
  //Poke to update price from Median
- function poke() public { |\label{line:poke}|
  //Handler function to receive price from Median
+ function prUpdt(uint wut,bool ok) handler { |\label{line:handlerosm}|
-  require(block.timestamp >= b + HOUR);
-  //peek current price
-  (uint wut,bool ok) = Median(median).peek();
   if (ok) {
-   cur = nxt;nxt = Feed(wut, true);
-   b = uint(block.timestamp-(block.timestamp % HOUR));
+   DPr.emit(wut,this.address).target(SigT).delay(HOUR);|\label{line:emitosm}|
    } }
  constructor(address median, ...) public {
   // handler prUpdt binds to signal Pr
+  prUpdt.bind(median,Median.Pr,0.1,false,SigR,SigM);|\label{line:bind1}|
   ... } }
- contract Spotter{  
-  //poke to update price from OSM
-  function poke(address osm) public { |\label{line:spotpoke}|
-   (bytes val, bool has) = OSM(osm).peek();
-   //calculate variable "spot" based on "val"
-   ...
-   vat.file(bytes(osm),"spot",spot);//file price to Vat |\label{line:file}|
-  } ... }
contract Vat {  // move here the body of Spotter contract
+  ...
+ address[] SigR; //allowed accounts |\label{line:SigR2}|
+ bytes[] SigM; //accessible methods |\label{line:SigM2}|
  //Handler function to receive price from OSM
+ function prUpdt(uint data,address osm) handler { |\label{line:handlervat}|
   //calculate variable "spot" based on "data"
    ...
+  file(bytes(osm),"spot",spot); //file price to Vat
  }
  // CDP Confiscation used for CDP liquidation
  function grab(...) external {...} |\label{line:grab}|
  // move function used for Dai join or exit.
  function move(...) external {...} |\label{line:move}|
  constructor(address osm0, address osm1, ...) public {
    // bind handlers to OSMs PriceFeed signal
+   prUpdt.bind(osm0,OSM.DPr,0.1,true,SigR,SigM);|\label{line:bind2}|
+   prUpdt.bind(osm1,OSM.DPr,0.1,true,SigR,SigM);|\label{line:bind3}|
    ...  } ... }
\end{lstlisting}


\noindent \textbf{MakerDAO components:} 
There are four contracts in Listing~\ref{lst:maker},
\sourcecode{Median}, \sourcecode{OSM}, \sourcecode{Spotter} and
\sourcecode{Vat}.  A \sourcecode{Median} contract is an oracle that maintains the
spot price of one digital asset type. A group of authorized maintainers send
batched transactions via the \sourcecode{feed} function to report the real
world spot price (line~\ref*{line:fd}). The function computes
the median of the reported price as the oracle spot price. The latest oracle price can be accessed via the \sourcecode{peek}
function (line~\ref*{line:pk}).

\sourcecode{OSM} is the acronym of Oracle Security Module. \sourcecode{OSM} delays the price
obtained from \sourcecode{Median} by an hour. Based on the MakerDAO
documentation, this is to allow an emergency committee to supervise the oracle
prices. In case of an emergency, the committee may pause the price feed and the
MakerDAO system via administrative interfaces (omitted in the code snippet).
\sourcecode{Spotter} is a contract that collects spot prices from multiple
\sourcecode{OSM} contracts for different kinds of digital asset collaterals.

\sourcecode{Vat} is the core vault engine which stores and tracks all the Dai
and collaterals. \sourcecode{Spotter} eventually invokes \sourcecode{Vat} to
file the price change (line~\ref*{line:file}). Note that all other modules rely on
\sourcecode{Vat} to implement the desired finance service for users. For
example, when a user joins the MakerDAO protocol to mint Dai with Ethereum or
other digital assets as collaterals, MakerDAO calls the function
\sourcecode{move} (line~\ref*{line:move}) in \sourcecode{Vat} to update the vault. When a
user submits a transaction to liquidate a position (i.e., sell the
collaterals), MakerDAO calls the function \sourcecode{grab} (line~\ref*{line:grab}) to
update. 

\begin{figure}[t]
\centering
\includegraphics[width=85mm,scale=0.5]{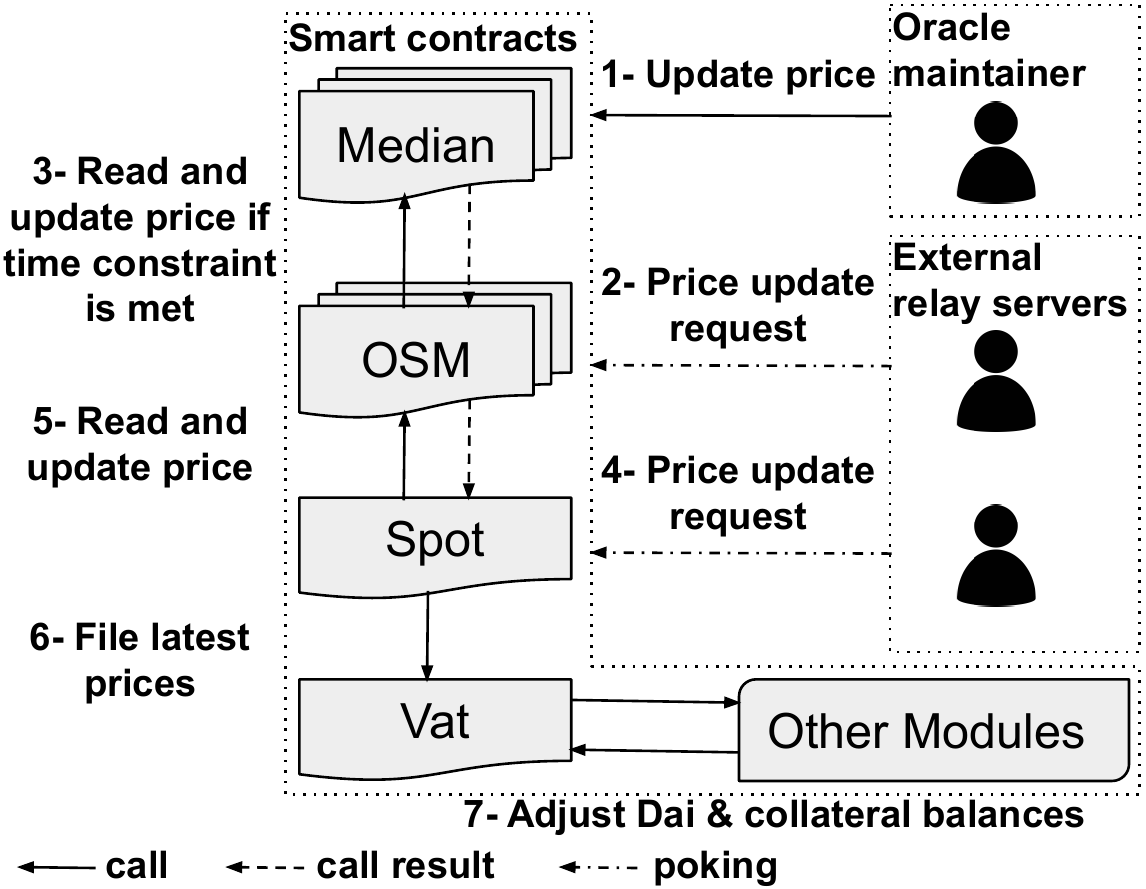}
\vspace{-8mm}
\caption{Overview of smart contracts and user interaction in MakerDAO.}
\label{fig:makerWithoutSig}
\vspace{-7mm}
\end{figure}

\noindent \textbf{Off-chain relay server poking:} 
It is not feasible to implement this price information propagation fully on
chain via function calls because the oracle maintainers and the MakerDAO
administrators are different groups of people. The asset price oracle service
might be used by many different DeFi contracts and the oracle maintainers are
not willing to pay excess transaction fee cost for the executions of MakerDAO.
Also, \sourcecode{OSM} needs to introduce one hour delay to the price feed,
which is not feasible to implement on-chain in Ethereum. 

MakerDAO therefore relies on off-chain relay servers to drive the price
information flow. Fig.~\ref{fig:makerWithoutSig} illustrates the interaction between off-chain servers and the contracts. The off-chain relay servers are expected to call \sourcecode{poke} in \sourcecode{OSM} (line~\ref*{line:poke}) for each digital asset every one hour to extract the price from the \sourcecode{Median} contracts. They are also expected to call
\sourcecode{poke} in \sourcecode{Spotter} (line~\ref*{line:spotpoke}) frequently to feed
the price information to the core engine. Function \sourcecode{poke} files
the price change into \sourcecode{Vat}, which will update its internal state to
change the behaviors of the implemented Dai join, exit, and liquidation functionalities accordingly.

\noindent \textbf{Security risks and loss:} 
The off-chain relay server design has significant security risks.
Knowing a reasonably accurate price for digital assets is
critical for the security of MakerDAO. If the off-chain relay server
fails or the network is congested so that poke transactions are not
processed in time, the core MakerDAO engine would operate
with outdated price information and make incorrect liquidation decisions.

On March 12 2020, the price of crypto currencies dropped significantly.
Meanwhile, the Ethereum network became overwhelmed with too many transactions.
Critical poke transactions containing price information were delayed, causing
the core MakerDAO engine to operate with stale prices for hours. As a result,
many MakerDAO users had their positions liquidated while getting much less
collateral back comparing to the amount they should have retained with the
correct prices. Even after the prices returned to the required level, the
collateral was auctioned because its price oracle failed to update price
feed. The root cause of this event is the market crash and the network
congestion, but the off-chain relay servers design exacerbated the delay during
the market collapse. The total financial loss of MakerDAO users during this
event is approximately \$4.5M~\cite{makerdao2020}.

\noindent \textbf{{\namelang} powered by {\name}:} We next show how to
implement the oracle components of MakerDAO in {\namelang}, our modified
Solidity language that can utilize special {\name} operations for an
event-driven execution model.

{\namelang} allows a contract to define its signal events via the
\sourcecode{signal} keyword (line~\ref*{line:signal}). The contract can emit
declared events during its execution together with event data, targeted
contracts to receive the events, and a delay in number of blocks
(line~\ref*{line:emit}). {\namelang} also introduces a new function modifier
\sourcecode{handler}. Functions declared with the keyword \sourcecode{handler}
are signal handler functions that can be attached to signal events
(lines~\ref*{line:handlerosm} and~\ref*{line:handlervat}). Once bound, handler
functions will be invoked as a separate signal transaction with the data
supplied by the event as parameters. 

Note that the third parameter of a \sourcecode{bind} statement is a positive
real number as the gas incentive provided by the contract. For example, $0.1$
in line~\ref*{line:bind1},~\ref*{line:bind2},~\ref*{line:bind3} denotes that
the contract is willing to pay a gas price $10\%$ higher than the average gas
price of other regular transactions in the same block computed by miners during
transactions packing. The fourth parameter of a \sourcecode{bind} statement is
either $\mytrue$ or $\myfalse$, indicating whether the generated signal
transactions from this handler function will lock the contract from
regular transactions or not. If it is $\mytrue$, then the contract locking
mechanism is enabled. Regular transactions will not be able to interact with
the contract if there is a pending signal transaction. 
The fifth and sixth parameters specify exceptions for the contract locking
mechanism, i.e., white-lists of addresses and methods that are allowed for
regular transactions during locking
(\sourcecode{SigR} and \sourcecode{SigM} at
lines~\ref*{line:SigR1} and~\ref*{line:SigR2}).

\noindent \textbf{Implementation of Median, OSM, and Vat in {\namelang}:}
To implement the desired price information flow, we define a signal event in
\sourcecode{Median} (line~\ref*{line:signal}) and a signal event in \sourcecode{OSM} (line~\ref*{line:signalosm}). Instead of passively waiting for other
contracts to call \sourcecode{peek}, \sourcecode{Median} emits a price feed
\sourcecode{Pr} event whenever there is a valid batch of oracle price
updates (line~\ref*{line:emit}). This event is automatically handled by the function in
\sourcecode{OSM} at line~\ref*{line:handlerosm}, which in turn emits a delayed price feed
\sourcecode{DPr} event with a delay of one hour. The emitted
\sourcecode{DPr} event will be eventually handled by the \sourcecode{prUpdt} function in \sourcecode{Vat} (line~\ref*{line:handlervat}) to file the new price information. This function replaces the original \sourcecode{poke} in \sourcecode{Spotter}. 
The contract \sourcecode{Spotter} is merged into \sourcecode{Vat}.

\noindent \textbf{Event-driven execution with {\name}:} Once deployed on
{\namechain}, the modified code in Listing~\ref{lst:maker} will enable the
desirable information to flow fully on-chain. 
When the oracle maintainers send a transaction to invoke the
function \sourcecode{feed} in \sourcecode{Median} (line~\ref*{line:fd}) to provide a
new price. The contract will emit the \sourcecode{Pr} event (line
15) with the calculated mean asset price as the event data. Because
\sourcecode{OSM} registered the function \sourcecode{prUpdt} (line~\ref*{line:handlerosm})
as the handler for \sourcecode{Pr} event, {\name} will
generate a signal transaction to invoke \sourcecode{prUpdt} with the corresponding event data as the parameter.  {\name} will execute the generated signal transaction automatically and asynchronously. 
The transaction will cascadingly emit the \sourcecode{DPr} signal event (lines~\ref*{line:emitosm}) with the asset price and an address id (which represents the digital asset type). This event will be emitted with a delay of one
hour. Because \sourcecode{Vat} registers the function \sourcecode{prUpdt} to handle the \sourcecode{DPr} event from
\sourcecode{OSM} contracts, after one hour delay, {\name} will generate and
execute a signal transaction to invoke \sourcecode{prUpdt} with the
asset price and the asset address as parameters automatically. \sourcecode{prUpdt} in \sourcecode{Vat} will finally file the price change at line~\ref*{line:file}.

\noindent \textbf{Contract event locking:}
The advantage of setting the handler function in \sourcecode{Vat} is to utilize the contract event lock mechanism in
{\name}. After the one-hour delay of the \sourcecode{DPr} event, {\name} will
lock the \sourcecode{Vat} contract from arbitrary regular transactions until the
signal transaction that invokes \sourcecode{prUpdt} is executed. This
prevents \sourcecode{Vat} from interacting with potentially malicious users before
it incorporates critical updates from the signal transaction, e.g. minting Dai
or liquidating collaterals with outdated digital asset prices. This lock
mechanism is lightweight because it only affects one contract. Other contracts
deployed on the blockchain are not affected. 

\noindent \textbf{Transaction fees:} Different from the Ethereum gas mechanism that transaction fees are always
paid by external users. The transactions fees of signal transactions in
{\namechain} are paid by the smart contracts who register the corresponding
handler functions. In our example, \sourcecode{OSM} and \sourcecode{Vat} must have sufficient native token balance to
cover the signal transaction fees. We believe this is a much smaller burden than
maintaining an off-chain relay server to send poke transactions.

\noindent \textbf{Advantages:} This example highlights the advantages of
{\name}. {\name} eliminates the dependency on the unreliable poking mechanism and off-chain relay servers. The execution model of {\name} guarantees that the core vault engine operates with the one-hour delayed prices
reported by the oracle maintainers. Listing~\ref{lst:maker} highlights the expressiveness of {\namelang}. Challenging features like time delays can be implemented in {\namelang} in a straightforward way.

\vspace{-1mm}
\section{{\name} Design}
\label{sec:design}

This section formalizes the design of {\name}. Similar to other smart contract
virtual machines like EVM~\cite{buterin2014ethereum}, {\name} contains two
layers, the virtual machine execution layer that dictates how {\name}
executes a transaction and the block processing layer that
dictates how the blockchain state evolves over multiple transactions in a
block. 
\vspace{-1mm}
\subsection{Core {\name} Language}

\begin{figure}
{\footnotesize
\setlength{\grammarindent}{7em}
\setlength{\grammarparsep}{0.15cm}
\begin{grammar}
<prog> ::= \texttt{program}  <inst>$^{*}$
    
<inst> ::=  <EVMinst> | <SIGinst> 
    
<SIGinst> ::= \texttt{createsignal} $\signame$ | \texttt{deletesignal} $\signame$ | \texttt{detach} $(\ameth, \signal)$ | \texttt{bind} $(\ameth, \signal, \gaspr, \blk, \sigRoles, \sigMeth)$  | \texttt{emit} $(\signal,\sigTarget,\delay)$ 
\end{grammar}}
\vspace{-3mm}
\caption{Core language signal related instructions syntax of {\name}. $a^{*}$ indicates zero or more occurrences of $a$.}
\label{fig:languageSyntax}
\vspace{-5mm}
\end{figure}

Fig.~\ref{fig:languageSyntax} lists the syntax of a simple programming language
used to formalize our approach. Our language extends the standard
EVM operations $\texttt{EVMinst}$ (e.g., $\texttt{load}$ and $\texttt{push}$)
with new operations $\texttt{SIGinst}$, e.g., $\texttt{createsignal}$ and
$\texttt{bind}$, to enable an event-driven execution model under
{\name}. For brevity, in Fig.~\ref{fig:languageSyntax} we omit standard EVM
operations $\texttt{EVMinst}$ since they are not necessary in understanding the
design of {\name}.  However, our implementation of {\name} actually supports
all standard EVM opcodes including arithmetic and inter-contract call
operations. 


In $\texttt{SIGinst}$,  $\texttt{createsignal}$ and $\texttt{deletesignal}$
opcodes handle the creation and deletion of a signal event with name $\signame$, 
respectively. The opcodes $\texttt{bind}$ and $\texttt{detach}$ allow
handler functions to listen and unlisten to signal events.

$\texttt{bind} (\ameth,
\signal, \gaspr, \blk, \sigRoles, \sigMeth)$  attaches a handler function
$\ameth$ to a signal event $\signal \in \mathbb{S}$ with a gas ratio $\gaspr$.
The signal transaction fee is computed by miners by calculating the average gas
price of regular transactions packed in the same block multiplied by $1+\gaspr$.
The last three parameters of $\texttt{bind}$ dictates how the locking mechanism
is enforced for the signal event: a boolean flag $\blk$ which when it is
$\myfalse$ regular transactions are allowed to execute. Otherwise,  when $\blk$
is $\mytrue$ only regular transactions initiated by accounts with addresses in
the array $\sigRoles$ and are executing functions in the array $\sigMeth$ are
allowed to execute. The contract is locked for other regular transactions until
the pending signal transactions finish. $\texttt{detach} (\ameth, \signal)$ detaches a handler function $\ameth$ from a
signal $\signal$. Finally, $\texttt{emit} (\signal,\sigTarget,\delay)$
emits signal events for signal $\signal$ with a delay of $\delay$.
$\sigTarget$ is an array of addresses that specifies the contracts that will be poked by the emitted signal. 
If $\sigTarget$ is not empty then $\signal$ pokes only contracts with addresses in $\sigTarget$. Otherwise, when
$\sigTarget$ is empty, $\signal$ pokes all contracts that contain handler functions attached to $\signal$. 
For each poked contract, a signal transaction that executes the corresponding handler function is created.



\begin{figure*}
\begin{mathpar}
\vspace{-2.5mm}
\inferrule{
    \texttt{createsignal}\ \signame \in\instrOf(\lsconf(\pcconf))\\
    \sighand[(\addr,\signame)] = \nil  \\
    \gasp := \priceOf(\texttt{createsignal})\\ 
    \sighand' := \sighand[(\addr,\signame) \mapsto \emptyset] \\
    \lsconf' := \lsconf[\pcconf\mapsto \mathsf{next}(\lsconf(\pcconf))]\\
}{
    (\gas, \_, (\aniden, \ameth, \addr, \lsconf) \circ \rtaskconf, \bbalance, \sighand,\_)\llll\texttt{createsignal}\ \signame \rrrr (\gas + \gasp, \_, (\aniden, \ameth, \addr, \lsconf') \circ \rtaskconf, \bbalance, \sighand',\_) 
} \and
\vspace{-1.5mm}
\inferrule{
    \texttt{deletesignal}\ \signame \in\instrOf(\lsconf(\pcconf))\\
    \sighand[(\addr,\signame)] \neq \nil  \\
    \gasp := \priceOf(\texttt{deletesignal})\\ 
    \sighand' := \sighand[(\addr,\signame) \mapsto \nil] \\
    \lsconf' := \lsconf[\pcconf\mapsto \mathsf{next}(\lsconf(\pcconf))]\\
}{
    (\gas, \_, (\aniden, \ameth, \addr, \lsconf) \circ \rtaskconf, \bbalance, \sighand,\_)\llll\texttt{deletesignal}\ \signame \rrrr (\gas + \gasp, \_, (\aniden, \ameth, \addr, \lsconf') \circ \rtaskconf, \bbalance, \sighand',\_) 
} \and
\vspace{-1.5mm}
\inferrule{
    \texttt{bind}\ (\ameth', \signal, \gaspr, \blk, \sigRoles, \sigMeth) \in\instrOf(\lsconf(\pcconf))\\
    \signal = (\addr',\signame) \\ 
    \sighand[\signal] \neq \nil  \\
    (\_, \addr, \_, \_, \_, \_) \not\in \sighand[\signal] \\
    \gasp := \priceOf(\texttt{bind})\\
    \sighand' := \sighand[\signal \mapsto (\ameth', \addr, \gaspr, \blk, \sigRoles, \sigMeth) \uplus \sighand[\signal]] \\
    \lsconf' := \lsconf[\pcconf\mapsto \mathsf{next}(\lsconf(\pcconf))]\\
}{
    (\gas, \_, (\aniden, \ameth, \addr, \lsconf) \circ \rtaskconf, \bbalance, \sighand,\_)\llll\texttt{bind}\ (\ameth', \signal, \gaspr, \blk, \sigRoles, \sigMeth) \rrrr (\gas + \gasp, \_, (\aniden, \ameth, \addr, \lsconf') \circ \rtaskconf, \bbalance, \sighand',\_) 
} \and
\vspace{-1.5mm}
\inferrule{
    \texttt{detach}\ (\ameth', \signal) \in\instrOf(\lsconf(\pcconf))\\
    \signal = (\addr',\signame) \\ 
    (\ameth', \addr, \_, \_, \_, \_) \in \sighand[\signal] \\
    \gasp := \priceOf(\texttt{detach})\\
    \sighand' := \sighand[\signal \mapsto \sighand[\signal] \setminus (\ameth', \addr, \_, \_, \_, \_)] \\
    \lsconf' := \lsconf[\pcconf\mapsto \mathsf{next}(\lsconf(\pcconf))]\\
}{
    (\gas, \_, (\aniden, \ameth, \addr, \lsconf) \circ \rtaskconf, \bbalance,\sighand,\_)\llll\texttt{detach}\ (\ameth', \signal) \rrrr (\gas + \gasp, \_, (\aniden, \ameth, \addr, \lsconf') \circ \rtaskconf, \bbalance,\sighand',\_) 
} \and
\vspace{-1.5mm}
\inferrule{
    \texttt{emit}\ (\signal,\sigTarget,\delay) \in\instrOf(\lsconf(\pcconf))\\
    \gasp := \priceOf(\texttt{emit})\\
    \esignal' := \mathsf{emitSig}(\signal,\esignal,\sighand,\sigTarget,\delay) \\
    \lsconf' := \lsconf[\pcconf\mapsto \mathsf{next}(\lsconf(\pcconf))]\\
}{
    (\gas, \_, (\aniden, \ameth, \addr, \lsconf) \circ \rtaskconf, \bbalance,\sighand,\esignal)\llll\texttt{emit}\ (\signal,\sigTarget,\delay) \rrrr (\gas+\gasp, \_, (\aniden, \ameth, \addr, \lsconf') \circ \rtaskconf, \bbalance,\sighand,\esignal') 
} 
\vspace{-1.5mm}
\end{mathpar}
\vspace{-6mm}
\caption{Program semantics in \name. For a function $f$, we use $f[a\mapsto b]$ to denote a function $g$ such that $g(c)=f(c)$ for all $c\neq a$ and $g(a)=b$. The function $\instrOf$ returns the instruction at some given control location while $\mathsf{next}$ gives the next instruction to execute. $\priceOf$ returns the gas price for a given opcode. We use $\circ$ to denote sequence concatenation.}\label{fig:sss}
\vspace{-3mm}
\end{figure*}

\vspace{-2mm}
\subsection{Operational Semantics}

A program configuration in {\name} is a tuple  $\mu = (\gas,\gsconf, $ \\ $
\rtaskconf, \bbalance,\sighand,\esignal)$ where $\mathit{gas}$ is the gas counter,
$\gsconf$ is composed of the persistent valuation of program variables,
$\rtaskconf$ is the call stack, $\bbalance$ is a mapping from addresses to the corresponding balances, $\sighand$ is a mapping that maps each signal to a set of handlers that are attached to this signal, and $\esignal$
is a set of emitted signal transactions. 

The activation frames in $\rtaskconf$ are represented using tuples $(\aniden,
\ameth, \addr, \lsconf)$ where $\aniden\in\mathbb{T}$ is a task (invocation)
identifier, $\ameth\in \mathbb{M}$ is a function name, $\addr \in \mathbb{A}$ is
the address of the contract that $\ameth$ belongs to, and $\lsconf$ is a
valuation of local variables, including a program counter. 

A signal identifier is a tuple $\signal = (\addr,\signame)$,
where $\addr \in \mathbb{A}$ is the address of the contract emitting the signal
 and $\signame$ is the signal name. $\sighand[\signal]$ is a set of tuples
$(\ameth, \addr, \gaspr,\blk, \sigRoles, \sigMeth)$ where $\ameth$ is the name of the handler function bound to 
the signal $\signal$ by the contract with
the address $\addr$, and $\gaspr$ is the gas ratio. 
The parameters $\blk$, $\sigRoles$, and $\sigMeth$ dictate the locking mechanism. 
$\blk$ is a boolean, $\sigRoles$ is an array of addresses, and $\sigMeth$ is an array of functions
signatures.  We assume that a signal $\signal = (\addr,\signame)$ exists only if $\sighand[\signal] \neq
\nil$ and if a signal is not attached to any method handler then we have $\sighand[\signal] = \emptyset$. 

The activation frames in the emitted signal transactions set $\esignal$ are
represented using tuples $(\anidenp, \signal, \ameth, \addr, \gaspr, \delay)$,
where $\anidenp\in\mathbb{SI}$ is a unique identifier of the signal
transaction, $\signal = (\addr',\signame)$ is the signal identifier, $\ameth$
is the handler method name, $\addr$ is the address of the contract containing
the handler method, $\gaspr$ is the gas ratio, and $\delay$ is the delay.

Fig.~\ref{fig:sss} presents the
small step semantics for {\name} signal related opcodes. The notation $\mu \llll \action \rrrr \mu'$
represents the state transition of $\mu$ to $\mu'$ after executing operation $\action$. 

A transition labeled by $\texttt{createsignal}\ \signame$ corresponds to the
creation of a new signal with the name $\signame$. The new signal is identified
with $\signal = (\addr,\signame)$ where is $\addr$ is the address of the
current contract. $\sighand[\signal]$ is set to the empty set. The gas counter
is incremented with the gas cost of the operation. A transition labeled by
$\texttt{deletesignal}\ \signame$ corresponds to the removal of the signal with
the name $\signame$. The signal is identified with $\signal = (\addr,\signame)$
where $\addr$ is the address of the current contract. $\sighand[\signal]$ is
set to $\nil$. 

A transition labeled by $\texttt{bind}\ (\ameth', \signal, \gaspr, \blk,
\sigRoles, \sigMeth)$ corresponds to attaching the signal $\signal$ to the
method $\ameth'$ and assigning $\gaspr$ as the gas ratio with locking parameters $\blk$, $\sigRoles$,
and $\sigMeth$. $\signal$ must exist, i.e., $\sighand[\signal] \neq \nil$.
Also, the current contract must not have bound another method to 
$\signal$, i.e.,  $(\_, \addr, \_, \_, \_) \not\in \sighand[\signal]$
where $\addr$ is the contract address. We then insert $(\ameth', \addr, \gaspr,
\blk, \sigRoles, \sigMeth)$ in the set $\sighand[\signal]$ to mark that the
contract $\addr$ assigned the method $\ameth'$ to $\signal$ with a gas ratio $\gaspr$. 
A transition labeled by $\texttt{detach}\ (\ameth',
\signal)$ corresponds to detaching handler $\ameth'$ from signal $\signal$.
The transition results in the removal of $(\ameth', \addr, \_, \_, \_, \_)$ from the set
$\sighand[\signal]$. 

Finally, a transition labeled by $\texttt{emit}\ (\signal,\sigTarget,\delay)$
corresponds to emitting a signal $\signal$ after a delay $\delay$. If
$\sigTarget$ is not empty $\signal$ pokes contracts
associated with addresses in $\sigTarget$. Otherwise, $\signal$ pokes
all contracts with attached handlers for $\signal$.
We use the function $\mathsf{emitSig}$ defined in Algorithm~\ref{algo1} to
update the emitted signal transactions set $\esignal$ with the newly emitted
signal transactions accordingly. 

\setlength{\textfloatsep}{10pt}
\begin{algorithm}[t]
  \caption{Signals to emit.}\label{algo1}
  \begin{algorithmic}[1]
  \Procedure{$\mathsf{emitSig}$}{$\signal$, $\esignal$, $\sighand$, $\sigTarget$, $\delay$}
  \State $\ \ \mathcal{Q} \leftarrow \esignal$; 
  \State \ \ \textbf{output} $\mathcal{Q}$;
  \State \ \ \textbf{for each}\ $(\ameth, \addr, \gaspr, \_, \_, \_) \in \sighand[\signal]$
  \State \ \ \ \ \ \ \ \textbf{if}\ $\sigTarget = \epsilon \text{ or }  \addr \in \sigTarget$
  \State $\ \ \ \ \ \ \ \ \ \ \ \ \anidenp \in \mathbb{SI}\ fresh$
  \State $\ \ \ \ \ \ \ \ \ \ \ \ \mathcal{Q}\leftarrow (\anidenp, \signal, \ameth, \addr, \gaspr, \delay) \uplus \mathcal{Q}$;
  \EndProcedure
  \end{algorithmic}
\end{algorithm} 

An execution of an externally invoked method $\ameth$ of a contract stored in the address $\addr$ is a sequence $\rho=\mu_0\llll\action_1\rrrr\mu_1\llll\action_2\rrrr\ldots\llll\action_n\rrrr\mu$ of transitions starting in the initial configuration $\mu_0 = (0, \gsconf, (\aniden, \ameth, \addr, \lsconf_0), \bbalance, \sighand,\epsilon)$ where $\lsconf_0 = \mathsf{locInit}(\gsconf,\ameth)$ represents the initial state of $\ameth$, and leading to a configuration $\mu = (\gas, \gsconf', \epsilon, \bbalance',$ $\sighand',\esignal)$ where the call stack is empty. We use the notation $\mu = \rho_{\ameth}(\mu_0)$  to say that the execution $\rho$ of $\ameth$ transforms $\mu_0$ to $\mu$.

Note that for brevity we omit the rules that check the current used
gas against the limit to terminate execution early. This is the standard
practice in Ethereum to address the termination problem and our {\name} implementation follows the same principle.

\vspace{-3mm}
\subsection{Transaction Execution}
\label{sec:design:tx-exec}


\noindent \textbf{Blockchain state: } A blockchain state in {\name} is a tuple $\sigma = (\height, \bsconf, \sighand,\ehsignal)$ where $\height$ is the current block height, $\bsconf$ maps each account address $\addr$ to  $\bsconf(\addr) = (\nonce,\bbbalance,\gsconf,\progcode)$ where $\nonce$ is the account nonce, $\bbbalance$ is the account balance, $\gsconf$ is the persistent valuation of contract variables, and $\progcode$ is the immutable contract code. 
$\ehsignal$ is a mapping that maps each block height to a set of emitted signal transactions pending until the block height is attained to be executed. For a block height $\height$, the activation frames in the emitted signal transactions set $\ehsignal[\height]$ are represented using tuples $(\anidenp, \signal, \ameth, \addr, \gaspr)$, where $\anidenp$ is the unique signal transaction identifier, $\signal$ is the signal identifier, $\ameth$ is the handler method name, $\addr$ is the address of the contract containing the handler method, and $\gaspr$ is the gas ratio. 


\begin{figure*}
\centering

\begin{mathpar}
\inferrule{
    (\nonce_s,\bbbalance_s,\gsconf_s,\progcode_s) = \bsconf(\addr_s) \\
    \bbbalance_s \geq \gaspp + \bbbalance_0 \\ 
    \addr_n \in \mathbb{A}\ fresh \\ 
    \forall\ \addr \in \mathbb{A}.\ \bsconf[\addr] = (\_,\bbbalance,\_,\_) \implies \bbalance_1[\addr] := \bbbalance \\
    \bbalance_2 :=\bbalance_1 [\addr_s \mapsto \bbalance_1[\addr_s] - \bbbalance_0;\ \addr_n \mapsto \bbalance_1[\addr_n] + \bbbalance_0] \\
    \gsconf_0 := \mathsf{init}(\addr_n) \\ 
    \lsconf_0 := \mathsf{locInit}(\gsconf_0,\ameth) \\
    (\gas, \gsconf, \epsilon, \bbalance, \sighand',\esignal) := \rho_{\ameth}(0, \gsconf_0, (\aniden, \ameth, \addr_n, \lsconf_0), \bbalance_2, \sighand,\epsilon) \\
    \gaspp \geq \gas \times \mathsf{unitGasPrice}(\height)\\ 
    \ehsignal' := \mathsf{signalsPartition}(\height, \ehsignal, \esignal)\\
    \bsconf_1 := \mathsf{updBal}(\bsconf,\bbalance) \\ 
    \bsconf_2 := \bsconf_1 [\addr_s \mapsto (\nonce_s + 1,\bbalance[\addr_s] - \gaspp,\gsconf_s,\progcode_s);\ \addr_n \mapsto (\nonce_s,\bbalance[\addr_n],\gsconf,\progcode) ]\\
}{
    (\height, \bsconf, \sighand,\ehsignal) \llll \texttt{createTx}\ (\addr_s,\nonce_s,\bbbalance_0,\ameth,\progcode,\gaspp) \rrrr^{t} (\height, \bsconf_2, \sighand',\ehsignal')
} \and
\inferrule{
    (\nonce_s,\bbbalance_s,\gsconf_s,\progcode_s) = \bsconf(\addr_s) \\
    (\nonce_t,\bbbalance_t,\gsconf_t,\progcode_t) = \bsconf(\addr_t) \\ 
    \forall\ \addr \in \mathbb{A}.\ \bsconf[\addr] = (\_,\bbbalance,\_,\_) \implies \bbalance_1[\addr] := \bbbalance \\
    \bbalance_2 :=\bbalance_1 [\addr_s \mapsto \bbalance_1[\addr_s] - \msgvalue;\ \addr_t \mapsto \bbalance_1[\addr_t] + \msgvalue] \\
    \bbbalance_s \geq \gaspp + \msgvalue \\
    \mathsf{lockPerm}(\height,\sighand,\ehsignal,\addr_t,\ameth,\addr_s) \\ 
    \lsconf_0 := \mathsf{locInit}(\gsconf_t,\ameth) \\
    (\gas, \gsconf, \epsilon, \bbalance,\sighand',\esignal) := \rho_{\ameth}(0, \gsconf_t, (\aniden, \ameth, \addr_t, \lsconf_0), \bbalance_2,\sighand,\epsilon) \\ 
    \gaspp \geq \gas \times \mathsf{unitGasPrice}(\height)\\ 
    \ehsignal' := \mathsf{signalsPartition}(\height, \ehsignal, \esignal)\\
    \bsconf_1 := \mathsf{updBal}(\bsconf,\bbalance) \\
    \bsconf_2 := \bsconf_1 [\addr_s \mapsto (\nonce_s + 1,\bbalance[\addr_s] - \gaspp,\gsconf_s,\progcode_s);\ \addr_t \mapsto (\nonce_t,\bbalance[\addr_t],\gsconf,\progcode_t) ]\\
}{
    (\height, \bsconf, \sighand,\ehsignal) \llll \texttt{regularTx}\ (\addr_s,\addr_t,\ameth,\msgvalue,\gaspp) \rrrr^{t} (\height, \bsconf_2, \sighand',\ehsignal')
} \and
\inferrule{
    \exists\ \height' \leq \height.\ \ehsignal[\height'] = (\anidenp, \signal,\ameth,\addr_t,\gasp) \uplus \textsf{q} \\
    (\nonce_t,\bbbalance_t,\gsconf_t,\progcode_t) = \bsconf(\addr_t) \\
    \gaspp := \gas \times \mathsf{unitSigGasPrice}(\height,\gaspr) \\  
    \bbbalance_t \geq \gaspp \\
    \forall\ \addr \in \mathbb{A}.\ \bsconf[\addr] = (\_,\bbbalance,\_,\_) \implies \bbalance_1[\addr] := \bbbalance \\ 
    \lsconf_0 := \mathsf{locInit}(\gsconf_t,\ameth) \\
    (\gas, \gsconf, \epsilon, \bbalance,\sighand',\esignal) := \rho_{\ameth}(0, \gsconf_t, (\aniden, \ameth, \addr_t, \lsconf_0), \bbalance_1,\sighand,\epsilon) \\ 
    \ehsignal_1 := \ehsignal[ \height' \mapsto \textsf{q}] \\
    \ehsignal_2 := \mathsf{signalsPartition}(\height, \ehsignal_1, \esignal)\\
    \bsconf_1 := \mathsf{updBal}(\bsconf,\bbalance) \\
    \bsconf_2 := \bsconf_1 [\addr_t \mapsto (\nonce_t + 1,\bbalance[\addr_t]  - \gaspp,\gsconf, \progcode_t) ]\\
}{
    (\height, \bsconf, \sighand,\ehsignal) \llll \texttt{signalTx}\ (\anidenp,\signal,\ameth,\addr_t,\gaspr) \rrrr^{t} (\height, \bsconf_2, \sighand',\ehsignal_2)
}
\end{mathpar}
\vspace{-5mm}
\caption{Transactions execution rules. 
$\mathsf{init}$ returns the initial state of a contract persistent variables. 
$\mathsf{unitGasPrice}$ returns the unit gas price at the block $\height$. 
$\mathsf{unitSigGasPrice}$ returns the average unit gas price payed by regular transactions 
packed in the block $\height$ multiplied by $1+\gaspr$. 
$\mathsf{updBal}(\bsconf,\bbalance)$ returns $\bsconf_1$ s.t. for every 
$\bsconf[\addr]=(\nonce,\bbbalance,\gsconf,\progcode)$, $\bsconf_1[\addr]=(\nonce,\bbalance[\addr],\gsconf,\progcode)$. }\label{fig:transact}
\vspace{-3mm}
\end{figure*}

\noindent \textbf{Transactions execution: } Fig.~\ref{fig:transact} presents 
the transaction execution rules for {\name}. We use $\sigma \llll \txaction \rrrr^{t} \sigma'$
to denote the blockchain state transition from $\sigma$ to $\sigma'$ after executing a transaction $\txaction$. 
In {\name}, we note three kinds of transactions: (1) create transaction which creates a new account and transfers balance, (2) regular transaction which executes a smart contract method and transfers balance, and (3) signal transaction which is a special transaction generated by signal emitted in previous blocks and does not transfer balance.

A transition labeled by $\texttt{createTx}\ (\addr_s,\nonce_s,\bbbalance_0,\ameth,\progcode,\gaspp)$ corresponds 
to an account creation transaction initiated by the account with the address
$\addr_s$ and nonce $\nonce_s$ transferring a balance $\bbbalance_0$ to the new
contract account. $\progcode$ parameter is for the code of a smart contract to
deploy in the new account and $\ameth$ is the name of constructor method in the
contract. $\gaspp$ is the gas price payed by the sender to execute the
transaction. The transition creates a fresh address $\addr_n$ for the new account
and ensures that the balance of the sender is sufficient to pay for the
transaction. Also, the transition includes an execution of the constructor
method $\ameth$, i.e., $\rho_{\ameth}$, over the newly created contract initial
state. The emitted signal transactions map $\ehsignal$ is updated with the
newly emitted signals during the execution of $\ameth$, i.e., $\esignal$ using
the function $\mathsf{sigPartition}$ that partitions emitted signal transactions based on the values of
the block height $\height$ and the delay period $\delay$ defined in
Algorithm~\ref{algo0}.
\setlength{\textfloatsep}{10pt}
\begin{algorithm}[t]
  \caption{Signal transactions partition.}\label{algo0}
  \begin{algorithmic}[1]
  \Procedure{$\mathsf{sigPartition}$}{$\height$, $\ehsignal$, $\esignal$}
  \State $\ \ \mathcal{Q} \leftarrow \ehsignal$; 
  \State \ \ \textbf{output} $\mathcal{Q}$;
  \State \ \ \textbf{for each}\ $(\anidenp, \signal, \ameth, \addr, \gaspr, \delay) \in \esignal$
  \State $\ \ \ \ \ \ \ \height' \leftarrow \height + \delay$ 
  \State $\ \ \ \ \ \ \ \mathcal{Q}\leftarrow\mathcal{Q}[\height'\mapsto(\anidenp,\signal,\ameth,\addr,\gaspr)\uplus\mathcal{Q}[\height']]$;
  \EndProcedure
  \end{algorithmic}
\end{algorithm} 

A transition labeled by $\texttt{regularTx}\
(\addr_s,\addr_t,\ameth,\msgvalue,\gaspp)$ corresponds to a regular transaction
initiated by the account with the address $\addr_s$ targeting the method
$\ameth$ in the contract associated with the address $\addr_t$ and
transferring a balance $\msgvalue$ to $\addr_t$. $\gaspp$ is the gas price payed
by the sender to execute the transaction. The transition ensures that the
balance of the sender is sufficient to pay for the transaction. To ensure that
the regular transaction is not executing a locked smart contract because of
pending signal transactions we define a boolean function
$\mathsf{lockPerm}$ which returns $\mytrue$ when the regular transaction
is allowed to execute and $\myfalse$ otherwise. 
\vspace{-3mm}
\begin{align*} 
    & \mathsf{lockPerm}(\height,\sighand,\ehsignal,\addr_t,\ameth,\addr_s) = \forall\ \signal\in\mathbb{S}.\ \\
    & (\ (\forall\ \height' \leq \height.\ \not\exists\ (\_,\signal,\_,\addr_t,\_,\_) \in \ehsignal[\height'])\ \vee\ (\exists !\ \\ 
    &  (\_, \addr_t, \_, \blk, \sigRoles, \sigMeth) \in \sighand[\signal]. \ \neg \blk \vee (\addr_s \in \sigRoles \wedge  \ameth \in \sigMeth)) \ ) 
  \vspace{-4mm}
\end{align*} 
$\exists !$ denotes unique existential quantifier. 
$\mathsf{lockPerm}$ returns $\mytrue$ when there are no pending signal
transactions ready to be executed and are associated with the contract $\addr_t$ that is targeted by the current regular transaction or for any pending signal transaction that can be executed we have that regular
transactions initiated by $\addr_s$ and are calling the method $\ameth$ are allowed to execute. If a contract detaches a handler from a signal while signal transactions are queued, regular transactions will be blocked until the signal transactions are executed ($\mathsf{lockPermission}$ returns $\myfalse$).

A transition labeled by $\texttt{signalTx}\ (\anidenp,\signal,\ameth,\addr_t,\gaspr)$ corresponds to
a signal transaction executing a method $\ameth$ in the contract with the address $\addr_t$. $\gaspr$ is the gas ratio. Similar to above, the transition ensures that the balance of $\addr_t$ is sufficient to
pay for the transaction. We remove the executed signal transaction
from the set of pending signal transactions at a block height less or equal to
the current block height, i.e., $\exists\ \height' \leq \height\
(\anidenp,\signal,\ameth,\addr_t,\gaspr) \in \ehsignal[\height']$.

\noindent \textbf{Block execution: } an execution of a new block of transactions $\ablock$ in {\name} is a sequence $\sigma_0 \llll \aincheight \rrrr^{h+1} \sigma_1::= \aincheight(\sigma_0)\llll \txaction_1 \rrrr^{t}\sigma_2\llll \txaction_2 \rrrr^{t}\ldots \llll \txaction_{n-1} \rrrr^{t}\sigma_n$ of transitions starting in the initial blockchain state $\sigma_0 = (\height, \bsconf_0, \sighand_0,\ehsignal_0)$ and leading to a blockchain state $\sigma_n = (\height + 1, \bsconf_n, \sighand_n,\ehsignal_n)$ where the transition $\sigma_0 \llll \aincheight \rrrr^{h+1} \sigma_1::= \aincheight(\sigma_0)$ simply increments the block height in the blockchain state $\sigma_0$ resulting in $\aincheight(\sigma_0) = (\height + 1, \bsconf_0, \sighand_0,\ehsignal_0)$ and $\txaction_1$, $\ldots$, $\txaction_{n-1}$ are the transitions of the transactions constituting the block $\ablock$, i.e., $\ablock = \{\txaction_1 \ldots  \txaction_{n-1} \}$.

\vspace{-2mm}
\section{Implementation}
\label{sec:implementation}

In this section we describe an implementation of an end-to-end blockchain
platform that supports {\name}. Our implementation consists of two main
components: {\namelang}, an extension of Solidity language together with a
compiler to support {\name} features, and {\namechain}, a blockchain client
based on the OpenEthereum client~\cite{openethereum}. As input, our
implementation accepts smart contracts written in {\namelang} and it then
deploys and executes them using {\namechain}. Our implementation supports all
EVM opcodes with the addition of {\name} opcodes for signal events presented in
Section~\ref{sec:design}.


\vspace{-2mm}
\subsection{{\namelang}}

Our {\namelang} language provides high-level language features for using
{\name} opcodes to create, bind, and emit signal events while maintaining
support for all Solidity language features. The {\namelang} compiler extends
the Solidity compiler~\cite{solc} with a new preprocessing parser that
transforms signal related statements to inlined assembly instructions that use
the corresponding {\name} opcodes. The resulting code is then complied to byte
code using the Solidity compiler which bypasses the inlined assembly code.
Thus, the outputted byte code includes signal related {\name} opcodes that were
introduced by the preprocessing parser and bypassed by the Solidity compiler.

The new high-level programming language features introduced by {\namelang} are the following:
\begin{itemize}[topsep=1pt]
\item \textbf{Signal declaration:} The statement \emph{signal S(parmType)} declares a signal event \emph{S} where
 \emph{parmType} is the list of data types associated with the event parameters. {\namelang} compiler translates the above statement into {\name} \emph{createsignal} opcode.

\item \textbf{Handler declaration:} using the keyword \emph{handler} a function is declared as a signal handler. For instance, \emph{function foo(...) handler \{ ... \}}  corresponds to defining  a handler function \emph{foo}. 

\item \textbf{Bind signal:} The statement \emph{foo.bind(addr, S, gr, blk, sigR, sigM)} binds the handler function $\emph{foo}$ to a signal event \emph{S} of a contract stored in the address \emph{addr}. \emph{gr} is the gas ratio. 
\emph{blk}, \emph{sigR}, and \emph{sigM} are the parameters for the locking mechanism to enforce. The above statement translates into {\name} \emph{bind} opcode.

\item \textbf{Detach signal:} The statement \emph{foo.detach(addr, S)} detaches the handler function $\emph{foo}$ from the  signal event \emph{S} of the contract stored in the address \emph{addr}. The above statement translates into {\name} \emph{detach} opcode.

\item \textbf{Emit signal:} The statement \emph{S.emit(parm).target(t).delay(d)} emits a declared signal event \emph{S} where \emph{parm} are the passed values of signal parameters, \emph{t} is an array of addresses of the recipients intended by the emitted signal, \emph{d} is a non-negative integer that specifies the number of blocks as the delay. The above statement translates into {\name} \emph{emit} opcode.
\end{itemize}
\vspace{-2mm}
\subsection{\namechain}
\label{sec:implementation:call}

Our {\namechain} extends Open\-Ether\-eum~\cite{openethereum} to support signal
transaction execution. In particular, it extends the Open\-Ether\-eum
EVM with the implementations of {\name} signal instructions. The execution of a
signal transaction in {\namechain} is similar to that of a regular transaction.
The main difference is that it starts at a handler function attached to emitted
signal rather than a user specified function. The blockchain state is augmented
with signal events related fields that we presented in
Section~\ref{sec:design}. Also, it maintains a list of all accounts with
pending signal transactions which provides an efficient method for miners to
retrieve pending signal transactions.  

In {\namechain}, a smart contract who registers a handler function will pay the
transaction fee cost of the associated signal transaction. If the smart
contract is unable to pay the fee of its signal transaction, the signal
transaction execution will stay in the pending transaction pool until the
contract obtains sufficient native tokens to cover the fee. Note that simple
native token transfer transactions that interact with the fallback method of
the contract are not affected by the contract locking mechanism by default.

\noindent \textbf{Enforcing contract locking: } In {\namechain} we extend the
transaction pool implementation in Open\-Ether\-eum. In particular, when
processing a block {\namechain} checks that for each: (1) regular transaction
whether the transaction is allowed under the locking mechanism or the pending
signal transaction queue for the contract is empty;  and (2) signal transaction
that the transaction is indeed the next transaction in the pending queue. If
these verification steps fail, the transaction will not be executed and will
have no effect. Miners who pack the transaction will receive no transaction fee
reward from the transaction. {\namechain} also enforces contract event locking
at the boundary of inter-contract calls. It checks whether the caller and the
called function are permitted under the locking mechanism or the pending signal
transaction list of the callee is empty. If the check fails, the whole
transaction will be reverted with no effect and the miner will receive no
transaction fee reward. Note that for the delegate call opcode, the check is
skipped because delegate calls do not change the state of the callee. 

Note that {\name} can handle denial-of-service attacks in which an attacker triggers a large
number of signal events to block regular transactions of a victim contract.
Such attacks are possible only when the victim contract listen to an event from
a contract deployed by the attacker in a way that blocks regular transactions.
Even so, the contract locking mechanism in {\name} allows the victim contract
to remove the locking partially by specifying a list of trusted addresses for
which regular transactions are not blocked.


\vspace{-3mm}
\subsection{Gas Mechanism}
\label{sec:impl:gas}

To incentivize miners to pack signal transactions as soon as possible, signal
transactions pay a higher gas price than regular transactions. In {\namechain},
we set the gas price of signal transactions to be \emph{the average gas price
of regular transactions in the block multiplied by one plus the gas incentive
ratio set during handler binding}. This mechanism ensures that the gas price of signal
transactions will always be competitive comparing to regular transactions. Miners
will always be incentivized to pack signal transactions even in the presence of
regular transactions with high fees. Because these regular transactions will
raise the gas price of signal transactions, thus miners earn the profits by
packing signal transactions instead of other regular transactions with low gas
fees.

To avoid situations where the average gas price of regular transactions cannot
be determined due to an overabundance of signal transactions, {\namechain} sets
a separate gas limit for signal transactions to reserve room for regular
transactions. {\namechain} limits the total gas consumption of signal
transactions in a block to be less than $\frac{1}{10}$ of the block gas limit.
We set this limit based on the current Ethereum ratio between contract
internal and regular transactions of $1:9$~\cite{etherscan}.

\vspace{-3mm}
\section{Evaluation}\label{sec:results}

In this section we outline an empirical study of {\name} using $23$ contracts from $13$ decentralized applications (DApps) on Ethereum using the implementation described in Section~\ref{sec:implementation}. Our empirical study is driven by the following questions:
\begin{enumerate}[topsep=1pt]
\item Can {\name} enable autonomous contracts? Specifically, can we
    implement popular smart contracts from Ethereum in {\name} to eliminate
    off-chain relay servers?
\item Is {\name} easy to use? What is the development effort of implementing smart contracts
    in {\name}?
\item What kinds of security risks are associated with off-chain relay server
    solutions? Can the contract event lock mechanism in {\name} eliminate these risks?
\end{enumerate}

\vspace{-2mm}
\subsection{Benchmark}

Our benchmark set contains $23$ contracts from $13$ DApps deployed on
Ethereum, including MakerDAO~\cite{makerdaowp}, Compound~\cite{compoundwp},
and Augur~\cite{augurwp}. 
Compound is a decentralized lending service platform where users could deposit one type of digital
asset as collateral to borrow another type of digital asset from its shared
pool. Augur is a decentralized prediction market where users bet on future 
result of real-world events. The supplementary material contains descriptions for the other DApps. 
In total, the $13$ DApps manage more than \$3B 
of digital assets on the Ethereum network, making up an essential part of the Ethereum economic ecosystem. 

\begin{table}
    \small
        \begin{tabular}{|m{6.1em}|m{5.5em}|m{1em}|m{1.35em}|m{1em}|m{1em}|m{1.35em}|m{1em}|}
        \hline
         &  & \multicolumn{2}{c|}{\textbf{Solidity}} & \multicolumn{4}{c|}{{\textbf{\namelang}}} \\
        \hline 
        \textbf{Application} & \textbf{Contract} & \textit{pf} & \textit{loc} & \textit{se} & \textit{hf} & \textit{loc'} & \textit{diff}\\
        \hline
        \multirow{3}{*}{MakerDAO} & Median & 1 & 175 & 1 & 0 & 177 & 2 \\\cline{2-8}
                  &OSM & 2 & 178 & 1 & 1 & 174 & 6\\\cline{2-8}
               &Spot\&Vat & 1 & 116 & 0 & 1 & 116 & 5\\\cline{2-8}
        \hline
        Compound & Timelock & 3 & 113 & 1 & 1 & 115 & 12\\
        \hline
        Augur & Universe & 1 & 711 & 1 & 1 & 727 & 15\\
        \hline
        MultiSigWallet &MultiSig & 1 & 256 & 1 & 1 & 270 & 19 \\
       \hline
       Historical Price Feed & HistoricalPrice & 1 & 564 & 1 & 1& 555 & 4 \\
       \hline
       Idex & Exchange& 1 & 1223 & 1 & 1 & 1231 & 8 \\
       \hline
       Aave & LendingPool & 1 & 355 & 1 & 1 & 359 & 4 \\
       \hline
       Metronome & Metronome & 1 & 125 & 1 & 1 & 133&8 \\
       \hline
       DAIHardFactory & DAIHard & 1 & 505 & 1 & 1 & 533 & 29\\
       \hline
       \multirow{3}{*}{KyberDxMarket} & PriceFeed & 1 & 201 & 1 & 0 & 203 & 2 \\\cline{2-8}
                                      & DSValue & 2 & 22 & 1 & 1 & 26 & 4\\\cline{2-8}
                                      & Medianizer & 1 & 195 & 0 & 1 & 197 & 2\\\cline{2-8}
       \hline
       \multirow{3}{*}{DutchExchange} & PriceFeed & 1 & 201 & 1 & 0 & 203 & 2 \\\cline{2-8}
                                        &DSValue & 2 & 22 & 1 & 1 & 26 & 4\\\cline{2-8}
                                        &Median & 1 & 2604  & 0 & 1 & 2606 & 2 \\\cline{2-8}
       \hline
       \multirow{3}{*}{PriceFeed} & PriceFeed & 1 & 201 & 1 & 0 & 203 & 2 \\\cline{2-8}
                                    &DSValue & 2 & 22 & 1 & 1 & 26 & 4\\\cline{2-8}
                                    &Median & 1 & 199 & 0 & 1 & 201 & 2 \\\cline{2-8}
       \hline
       \multirow{3}{*}{OracleInterface} & PriceFeed & 1 & 201 & 1 & 0 & 203 & 2 \\\cline{2-8}
                                        &DSValue & 2 & 22 & 1 & 1 & 26 & 4\\\cline{2-8}
                                    & Medianizer & 1 & 483 & 0 & 1 & 485 & 2 \\\cline{2-8}
       \hline
        \end{tabular}
    \caption{Empirical results. Characteristics of applications: lines of code (\textit{loc}) and number of poke functions (\textit{pf}). Characteristics of applications using signal events: number of signal events (\textit{se}), number of handler functions (\textit{hf}),  lines of code (\textit{loc'}), and lines of code difference (\textit{diff}).}            
    \label{table:sbc}
    \vspace{-5mm}
    \end{table}
Table~\ref{table:sbc} outlines some characteristics of our benchmark. Each DApp contains one or more contracts that are interdependent with each other but are maintained by different groups of users. 
In our evaluation, we focus on $23$ contracts (column \emph{Contract}) that are relying on off-chain relay
servers to invoke poke functions. 

\vspace{-2mm}
\subsection{Evaluation}

\noindent \textbf{Contracts in {\namelang}:} We reimplemented the benchmark contracts in {\namelang} with the goal of eliminating off-chain relay servers from the picture. Because {\namelang} is compatible with
Solidity, our modification is merely replacing the poke functions with
{\namelang} signal event statements. We are able to reimplement all $23$
contracts in {\namelang}. Columns \emph{se} and \emph{hf} in Table~\ref{table:sbc} present the number of declared signal events and the number of declared handler functions in the new contract implementation, respectively. 
We note that the new code is typically simpler than the original contract code while the reimplementation effort is moderate. For instance, Column \emph{diff} presents the number of lines difference between the new and old code. 

\noindent \textbf{Validation and analysis:} We deployed the new contracts implementation together with the remaining components of the $13$ applications in our {\namechain} platform. We validated that all three benchmark applications function correctly with the new contract code. 
We select the three applications MakerDAO, Compound, and Augur as case studies and we report our experience in analyzing the code of the new contracts. 

\vspace{-2mm}
\subsubsection{Case Study: MakerDAO}
As described in Section~\ref{sec:example}, we 
reimplemented the three contracts in the oracle components of
MakerDAO, eliminating the dependency on off-chain relay servers. The new
contracts powered by {\name} ensure that the core MakerDAO engine will always
operate with the updated price of its underlying digital asset. 
In addition, before the MakerDAO contract processes these price feed signal transactions, 
by lock mechanism, other transactions like liquidation and bidding will be put on-hold 
to avoid them operating with stale price information.

\begin{lstlisting}[caption={Code snippets of Timelock.sol in Compound}, label={lst:timelock}, language=Solidity,float,floatplacement=H, ]
contract Timelock {
  struct LockedTx {//Information of a transaction
    address target;uint txvalue;string signature;bytes datain;uint user_delay;bool ready;}
  // Transaction queue
  mapping (bytes=>LockedTx) private queuedTx;
  //max delay when poking executeTransaction
- uint public constant GRACE_PERIOD = 14 days;
+ uint public delay;// Base delay to execute
+ address[] SigR; //allowed accounts
+ bytes[] SigM; //accessible methods
+ signal TimeUp(bytes); //signal event
+ address[] SigT; //target handler addresses
  function queueTransaction(address target,uint txval,
    string memory signature,bytes memory data,uint userDelay)  
     public returns (bytes) {
    // Hash transaction parameters as index
    bytes txHash=keccak256(abi.encode(target,txval,signature,data,userDelay));
    //Replace timestamp with offset delay
-   require(eta>=getBlockTimestamp().add(delay));      
+   require(userDelay >= delay);
    // Push the new transaction to queuedTx
    queuedTx[txHash]=LockedTx(target,txval,signature,data,userDelay,true);
    // Emit signal event with txHash to itself
+   SigT=[this.address];
+   TimesUp.emit(txHash).target(SigT).delay(userDelay);
  // handler for executing tx
-  function executeTx(bytes txHash) public {
+  function executeTx(bytes txHash) handler {
     //remove check that sender is admin  
-    require(msg.sender == admin); 
     //remove checks that tx is in the correct period
-    require(getBlockTimestamp() >= eta); 
-    require(getBlockTimestamp() <= eta + GRACE_PERIOD);
      // use queuedTx[txHash] to do function call
      ... }
  constructor(address admin_, uint del,address[] roles_)public{
   //signal initializations
+  SigR = roles_; 
+  SigM = [abi.encode("queueTransaction(address,uint,
+           string,bytes,uint)"), ... ];
   //bind handler to signal
+  executeTx.bind(this.address,TimeUp,0.1,true,SigR,SigM);  
  } ... }
\end{lstlisting}
\vspace{-2mm}
\subsubsection{Case Study: Compound}
Compound protocol is a decentralized market to lend or borrow assets.
Users communicate with the Compound contracts to supply, withdraw, borrow
and repay assets. In the protocol, the key parameters of the market, such as
interest rate, risk model and underlying price, are managed by a
decentralized government body in Compound~\cite{compoundwp}. Any proposals of
updating the parameters must be voted for approval and queued in
\sourcecode{Timelock} contract. To give market participants time to react for any
parameter changes, the queued proposals will be executed in the required delay
specified in the proposals. 

We reimplemented the \sourcecode{Timelock} contract in {\namelang}
and validated it with our {\namechain} platform.
Listing~\ref{lst:timelock} presents the simplified code snippet of
\sourcecode{Timelock}. Lines preceded with ``-" and ``+" signs are changes we made to the code.

In the original Compound design, contract's Governor calls \sourcecode{queueTransaction} (line 13) to queue proposals. But to implement the desired delay,
Governor or an off-chain relay server needs to send another poke transaction to
call \sourcecode{executeTx} (line 27) right after the completion the delay period and before the proposal expiration time. \sourcecode{executeTx} first checks whether the proposal is queued and if the current timestamp falls in the
required interval (lines 32-33). If so, the proposal is executed. This design
is not desirable because the Governor or the off-chain relay server may fail and/or the poke transaction may not be packed in time if the blockchain network is congested.

In {\namelang}, we eliminate the dependency on poke transactions with the
signal event mechanism. We define a new signal event called \sourcecode{TimeUp}
at line 11. The \sourcecode{queueTransaction} function will emit this event
with the specified delay (line 25). Instead of being a public
function, the \sourcecode{executeTx} function is declared as an
event handler (line 28). When the contract is constructed, we 
bind \sourcecode{executeTx} as the handler for \sourcecode{TimeUp}
events in the same contract. Thus, \sourcecode{executeTx}
will be automatically executed once the specified delay time (measured in the
number of blocks) has past after each \sourcecode{TimeUp} event. Many conditional
checks in \sourcecode{executeTx} are removed because they are guaranteed by the execution engine of {\name}.

We deployed the modified \sourcecode{Timelock} contract together with the rest
of Compound smart contracts in {\namechain}. Our results show that
the deployed Compound application in {\namechain} can operate successfully and
there is no need to run off-chain server to poke the \sourcecode{executeTx} function anymore.

\vspace{-2mm}
\subsubsection{Case Study: Augur}
Augur is a decentralized prediction market where users can bet on future 
real-world outcomes~\cite{augurwp}. Augur markets follow four stages:
creation of the market, trading, reporting outcomes and prediction settlement.
During reporting stage, one of the reporters report results of a real-world
event by staking REP tokens on one of the outcomes of the market. If a Dispute
Bond is reached on an alternative outcome, the Tentative Winning Outcome
changes to the new alternative outcome. This process happens every $24$ hours.

The contract that implements this process in Augur is \sourcecode{Universe}.
The design of the contract is that every $24$ hours, the \sourcecode{sweepInterest} function has to be executed so that the contract can finalize the last round results and can start a new round if required.
Because this functionality is not feasible to implement in EVM, Augur instead
relies on external users to periodically poke \sourcecode{sweepInterest}.

\begin{lstlisting}[language=Solidity, caption={Code snippets of Universe.sol in Augur},label={lst:universe}, float,floatplacement=H,]
contract Universe is IUniverse {
    //define MIN_RATE as min of exec period
    //daily block rate of the network
    uint public dBlkR;
+   signal DSig(); //daily signal event
+   address[] SigT; //target handler addresses
+   address[] SigR; //allowed accounts
+   bytes[] SigM; //accessible methods
    //handler function
+   function Update() handler { 
+    sweepInterest();
     //signal emission
+    SigTargets = [this.address];
+    DSig.emit(this.address).target(SigT).delay(dBlkR); }
    function sweepInterest() public returns (bool) {...}
+   function setBlkRate(uint _dBlkR) public {
+    require(_dBlkR >= MIN_RATE);
+    dBlkR = _dBlkR; }
+   function start_emit() public {
     //emit a signal immediately 
+    SigT = [this.address];
+    DSig.emit().target(SigT).delay(0); }
    constructor(Augur _augur, ..., uint _dBlkR) public {
+    dBlkR = _dBlkR; //set handler execution period
     //update SigRoles and SigMethods
+    ...
     //bind handler to signal
+    Update.bind(this.address,DSig,0.1,true,SigR,SigM);
      ... } ... }
\end{lstlisting}
Listing~\ref{lst:universe} presents the code snippet of our
\sourcecode{Universe} contract implementation in {\namelang}. We define a
delay signal event called \sourcecode{DSig} at line 5 and we define a handler
function called \sourcecode{Update} for this event (lines 10-14). The handler
function calls \sourcecode{sweepInterest}. It also recursively
emits the \sourcecode{DSig} event with a delay of one day (in the number
of blocks) at the end (line 14). Therefore once the first
\sourcecode{DSig} event is triggered (via the
\sourcecode{start\_emit} function), the \sourcecode{Update} function will be
automatically invoked once every 24 hours. 

\begin{figure}[t]
    \centering
    \includegraphics[width=3in]{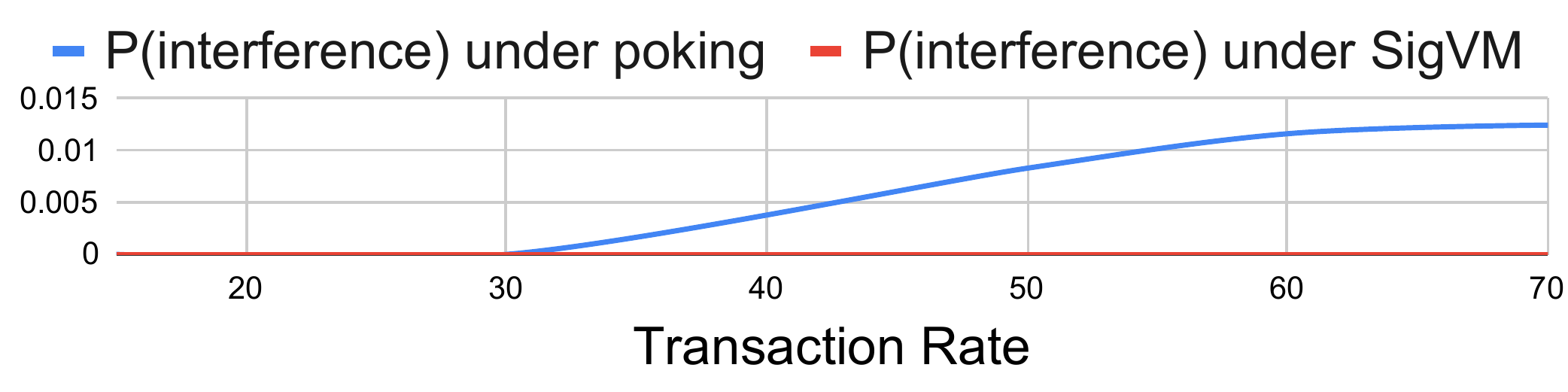}
    \vspace{-4mm}
    \caption{Malicious Hijacking Manipulation Rate Line Chart}
    \label{fig:erlc}
    \vspace{-2mm}
\end{figure}

We deployed the new \sourcecode{Universe} contract together with the rest of
contracts in Augur in our {\namechain} platform. The deployed version of Augur
operates successfully and the \sourcecode{sweepInterest} function is
automatically called every 24 hours as expected. This example highlights the
expressiveness of {\namelang} powered by {\name}. This recursive event handler
pattern enables developers to create customized infinite timer ticks
conveniently. The desired handler function will be automatically and
periodically invoked as signal transactions by the execution engine of {\name},
as long as the contract has enough native tokens to pay for the transaction fee
of the recurring transactions.

\vspace{-2mm}
\subsection{Eliminate Risks of Off-chain Relay Server}

We describe an experiment to quantitativaly study malicious hijacking manipulations caused
by relying on off-chain relay server to drive execution. We also evaluate whether {\name} can eliminate those associated risks.

\noindent \textbf{Malicious hijacking manipulation:} 
Using off-chain relay servers to simulate event-driven execution models may
make the contract vulnerable to malicious hijacking manipulations. Miners may pack an
interference transaction before a poke transaction for an event and therefore
the contract will execute the interference transaction in an incorrect state.

We setup an experiment with the MakerDAO contracts to quantitatively illustrate this problem. We start an Ethereum network with a block generation rate of one block per second. We developed a random \emph{transactions
generator} to generate three types of transactions, 1) simple transfer
transactions that do not interact with MakerDAO, 2) MakerDAO transactions, and
3) poke transactions that attempt to update price for MakerDAO. In the
experiment, the ratio of the three kinds of transactions is $10:4:1$, respectively. The gas
price for the three transactions follows the normal distribution with an average: $15$,
a standard deviation: $2$, and a maximum: $50$.

Fig.~\ref{fig:erlc} presents, using an off-chain relay server, the chance of manipulating other MakerDAO transactions to be executed ahead of a poke transaction as we increase the transaction throughput.
Hijacking manipulation occurs starting from a transaction rate $\ge 30$ TPS (Transactions Per Second) and goes as high
as $0.0125$ when the throughput is $70$ TPS. We repeat the experiments with the new MakerDAO contracts in our {\namechain} platform by replacing poke transactions with the automatic signal event mechanism. The hijacking manipulation never occurs because of our novel contract event lock mechanism. The lock mechanism has minimum impact on the performance because the miners are incentivized to pack signal transactions as soon as possible.

\if false
Under Node A, after initializing the contract \verb|Timelock|, the transaction, which
is a function call of \sourcecode{setDelay()}, is queued. After 100s, which is equivalent to 
100 relative block height,
\sourcecode{setDelay()} will be called by executing \sourcecode{executeTx()}. After
consensus syncing, under Node B, function \sourcecode{getDelay()} is locally called to
check whether \sourcecode{Delay} is updated or not. 
With the poking mechanism the consensus delay between node A \& B is 2 relative block heights on average.
The delay of relay server's starting to poke is 1 relative block height on average.\\
During experiments, \sourcecode{queueTransaction()} is poked 
at relative block height 0. Under poking mechanism, relay server pokes \sourcecode{executeTx()} 
at height 100. \sourcecode{Delay} is updated at height 103 as expected. $103-100=3$ relative block heights 
are the timing skew. Under {\name}, however, \sourcecode{Delay} is updated at height 100 and $100-100=0$ timing skew occurs. 
The reason is that after emitting the signal event in \sourcecode{queueTransaction()}, the
\verb|account_handler_queue| is synced to Node B. Then after 100 relative block heights, 
Node B executes the handler function locally. In practice, network congestion can cause even greater 
delays among relay servers. The timing skew of {\name} handler function execution is smaller 
than poking mechanism.
\fi

\vspace{-1mm}
\section{Related Work}\label{sec:relatedwork}

\noindent \textbf{Event-driven proxy services:} On Ethereum, there are a number
of proposals involving proxy service of function executions in respondence to
event emission. EventWarden~\cite{li2020eventwarden} is a proxy service for a user to create a smart contract specifying events to listen, the function handler to call when an event occurs, and the service fee
for the executor. Executors of the system monitor the event log, invoke the
specified function, and earn a service fee in return. Similarly,
Ethbase~\cite{ethbase} provides a registry smart contract where users can
deposit and subscribe to an automatic function invocation service. When the
specified event is seen in the log, miners invoke the callback function
accompanied by a proof of the event emission to the registry. Both EventWarden and Ethbase are variants of off-chain relay server solutions, although they provide incentives on chain via smart contracts so that
any external user could act as an off-chain relay server to send poke
transactions to claim the reward. 
They share similar weakness as the off-chain relay server
solution, i.e., the reliance on external users to send poke transactions timely
and their willingness to pay high enough fees for these poke transactions to be
packed timely. Different from these approaches, our solution modifies the
blockchain system to enable native support of an event-driven execution.
{\name} brings the whole execution process on-chain and completely eliminates the dependency on off-chain relay actors.


\noindent \textbf{Smart contracts programming: } In addition to Solidity, a number of 
new languages have been designed for smart contracts programming, including, Vyper~\cite{vyper}, Simplicity~\cite{simplicity}, Liquidity~\cite{liquidity}, Sophia~\cite{sophia}, Move~\cite{blackshear2020move}, DeepSEA~\cite{soberg2019deepsea}, and Obsidian~\cite{coblenz2019obsidian}. Unlike {\name}, the above languages do not support an event-driven execution model. Interestingly, because the event-driven execution model is highly demanded and many smart contracts are unsecurely simulating such an execution model via off-chain relay servers, {\name} also improves the security of these contracts by eliminating the need of such insecure practice. Also the aforementioned languages either introduces new syntaxes that leads to new learning curve for developers or suppress expressiveness in order to gain security guarantees.  On the other hand, {\namelang} is a practical extension on Solidity with minor changes in the syntax while improving security and usability of the language.


Recent smart contract languages such as RhoLang~\cite{rholang},
Scilla~\cite{sergey2019safer}, and Nomos~\cite{das2021csf} replace inter-contract function calls with message-passing mechanisms to eliminate bugs like re-entrance. Although those message-passing mechanisms are
also asynchronous, they behave otherwise like function calls and cannot implement the desired event-listener model to eliminate off-chain relay servers.  In fact, it would be difficult to implement many DeFi contracts like MakerDAO in those languages, because these contracts depend on synchronous inter-contract function calls. Making a function call asynchronous via message-passing may make such a contract vulnerable to malicious manipulation attacks that front-run the asynchronous
massage-passing transaction. 

General purposes programming languages such as JavaScript support event-driven asynchronous programming. However, they are not commonly used to write smart contracts. 

\noindent \textbf{Security and correctness validation:}
Another path taken by many to improve the security of smart contract programming is through static verification and runtime validation tools. There is a rich body of literature on detecting vulnerabilities in smart contracts
through static analysis and modular verification~\cite{Kalra2018ZEUSAS, luu2016oyente,
nikolic2018finding,bhargavan2016formalverification, verx, kevm}. For instance, ith ,
Oyente~\cite{luu2016oyente} uses symbolic executions to verify smart contracts against various attacks including re-entrance  and mishandled exception attacks.  Verx~\cite{verx} uses delayed abstraction to detect and verify temporal safety
properties automatically. Solythesis~\cite{li2019securing} inserts runtime checks to enforce customized validation rules. 
KEVM~\cite{kevm} defines a formal semantics of EVM in $\mathbb{K}$ and verifies smart contracts against user-defined specifications. These techniques enhance security of smart contract programming mostly by preventing unintended behaviors in smart contract codes and cannot be used to eliminate the dependency on off-chain relay servers.

\noindent \textbf{Scheduled transaction execution:} Bitcoin supports a native
mechanism called timelocks~\cite{antonopoulos2018bitcoin} to delay transaction
execution. The transaction-level timelock feature can be utilized by specifying an 
execution delay with the nLocktime field. Additional timelock features, Check Lock Time Verify (CLTV) and Check Sequence Verify
(CSV), were introduced later to the scripting language. CLTV limits the
availability of the associated Unspent Transaction Output (UTXO) until a
certain age. CSV utilizes the the value of nSequence, a transaction field, to
prevent mining of a transaction until the time limit specified for the UTXO is
met. These features are proven to be useful in layer 2 designs, such as
state channels and the Lightning Network. The timelock features provides a
convenient way to delay transaction execution on Bitcoin. However, with the
lack of programmability on Bitcoin network and the difficulty of generalizing
such a design to non-UTXO networks, the usability of such a design is limited.
Another implementation of delayed or periodic transaction is to setup relay
servers through the network client as we have discussed before this is not desirable. 



\vspace{-1mm}
\section{Conclusion}\label{sec:conclusion}

As smart contracts become more complicated and inter-dependent with each other,
the event-driven execution model is an increasingly demanded feature.
Unsatisfied by current blockchain virtual machines, smart contracts start to
use unreliable mechanisms such as off-chain relay servers to insecurely
simulate event-driven execution. {\name} provides the first blockchain virtual
machine with an integrated solution to natively enable event-driven execution
models on-chain. It paves the way for developers to build fully autonomous
and robust smart contracts in future.

\bibliography{paper}



\appendix
\section*{Appendix}
\renewcommand{\thesubsection}{\Alph{subsection}}
\setcounter{table}{0}
\renewcommand{\thetable}{A\arabic{table}}
\subsection{Applications Descriptions}
\label{subsec:Applications}
\noindent \textbf{MultiSigWallet With TimeLock~\cite{MultiSigWalletWithTimeLock}:} 
The Smart Contract program serves to allow multiple parties to agree on transactions before execution. 
Once all of the parties confirm one transaction, the transaction will be executed by anyone in a required delay.
In most cases, a relay server is responsible to poke to execute after the delay.
In order to eliminate the need of relay server, 
we reimplemented the contracts in SigSolid and the transaction is executed in the form of signal transaction.
In function \sourcecode{confirmTransaction()}, a signal event with transaction ID is emitted once all of the owners confirm.
In the contract \sourcecode{UserExecuteTransaction} of a user, the handler function \sourcecode{multiSigWalletExecuation()} binds with the signal.
After a required delay, the signal transaction of the user is executed and function \sourcecode{executeTransaction()} is invoked to execute the determined transaction.

\noindent \textbf{Historical Price Feed~\cite{HistoricalPriceFeed}:}
The contract is used to store updated historical price data from an off-chain oracle.
The function \sourcecode{poke()} should be manually invoked every 24 hours in order to update prices periodically.
In SigSolid, a signal event is declared without including any parameters. 
\sourcecode{poke()} is the corresponding handler function which emits the signal event inside.
In this way, \sourcecode{poke()} signal transaction is always executed periodically to update the price data.

\noindent \textbf{Idex~\cite{idex}:}
Idex is a platform which provides ERC-20 Token trading on Ethereum. 
Function \sourcecode{executeTrade()} in \sourcecode{Exchange} contract is responsible for settling trade orders.
Once the chain propagation period has elapsed, \sourcecode{executeTrade()} invalidate all trade orders whose timestamp is lower than the one provided.
In SigSolid, similar with Historical Price Feed, 
by setting empty signal emission in a handler function, 
order invalidations are executed every chain propagation period.

\noindent \textbf{Aave~\cite{aave}:}
Aave protocol is a decentralized lending pools protocol on Ethereum. 
Function \sourcecode{liquidationCall()} will be invoked by relay servers
to check and liquidate undercollateralized position.
There is a price update from an price oracle under such manual checking.
In order to mitigate the unexpected delay caused by relay server poking,
SigSolid introduces a periodical liquidation check mechanism similar with 
Historical Price Feed.
In contract \sourcecode{LendingPoolLiquidationManager} of Aave,
by setting empty signal emission in a handler function, the price is updated from oracle periodically.
In the contract of users, the handler function binding with the empty signal will invoke
\sourcecode{liquidationCall()} to check regularly.

\noindent \textbf{Metronome~\cite{Metronome}:}
Metronome is a DeFi trading market for a type of ERC-20 token on Ethereum.
In contract \sourcecode{Metronome}, 
function \sourcecode{poke(address a)} is used to create a reward from the funds of 
account \sourcecode{a} who has not idled in the last 10 minutes. 
The user who invokes \sourcecode{invest()} in the last 10 minutes is regarded 
as being under Not-Idle status.
In SigSolid, \sourcecode{poke()} is set as a handler function.
In \sourcecode{invest()},
First detach \sourcecode{poke()} with a old signal event if binding before.
Then \sourcecode{poke()} is re-bound with the signal with the sender address.
Then the signal is emitted with the required delay.
The purpose of re-binding action is to update the handler function execution timestamp
if the user keeps non-idle with the required delay.
Then function \sourcecode{poke()} will be executed without the function call from reply servers.

\noindent \textbf{DAIHardFactory~\cite{DAIHardFactory}:}
Contract DAIHardFactory provides a DAI trading platform. One trade process has five sequential phases. 
A fixed time interval called \sourcecode{"autoabortInterval"} is defined. 
During each phase, once \sourcecode{"autoabortInterval"} has passed, function \sourcecode{abort()}
is executed by anyone to abort from the phase. Otherwise, the next phase will be executed.
In SigSolid, at the beginning of each phase execution, 
a signal event with current phase type is emitted. 
After a delay of \sourcecode{"autoabortInterval"},
the corresponding handler function first checks whether the current phase has changed or not.
Next it aborts from the phase if the phase is not changed.

\noindent \textbf{Contracts similar with MakerDAO:}
KyberDxMarketMaker~\cite{KyberDxMarketMaker}, DutchExchange~\cite{DutchExchange},
PriceFeed~\cite{PriceFeed} and OracleInterface~\cite{PriceOracleInterface}
shown in the last four rows of Table~\ref{table:sbc} provide price oracle service similar with MakerDAO.
Users should use relay servers to manually poke to retrieve the latest price periodically. 
SigSolid mitigates the unexpected delay caused by relay servers. The price oracle emits
the price signal in a single transaction.
All of the handlers of users are executed after a required delay.

\end{document}